\documentstyle[12pt]{article}
\title{Neutrino Decay in the Doublet Majoron Model\\~\vspace*{3.0em}}
\author{Hisashi Kikuchi and Ernest Ma\\
{}~\vspace*{0.2em}\\
{\normalsize \sl Department of Physics}\\
{\normalsize \sl University of California, Riverside}\\
{\normalsize \sl Riverside, CA 92521} \\
{}~\vspace*{2.0em}
}
\date{UCRHEP-T126\\May 1994\\}
\newcommand{\bea}{\begin{eqnarray}}
\newcommand{\eea}{\end{eqnarray}}
\newcommand{\be}{\begin{equation}}
\newcommand{\ee}{\end{equation}}

\begin{document}
\maketitle
\thispagestyle{empty}

\vfil

\begin{abstract}
A new Majoron model is presented within the framework of the seesaw mechanism.
Its Higgs sector consists of only doublet representations
and the lepton-number violation takes place at the same scale of
the electroweak symmetry breaking.
This model is different from  the singlet-
or triplet-Majoron model in several respects:
it is free from the \(\rho\)-parameter constraint and it provides
moderately fast neutrino decay, but
the constraint from the stellar cooling of red giants
is satisfied only with an imposed approximate symmetry.
A \(\tau\) neutrino as heavy as 10 MeV is possible in this model
despite various cosmological and astrophysical constraints.
\end{abstract}

\vfil

\pagebreak
\baselineskip=24pt

\paragraph{1. Introduction}~

We present  in this paper a new  Majoron model.
Its Higgs sector consists only of doublet representations under the standard
electroweak gauge group G  \(\equiv\)  SU(2)\(\times\)U(1)\(_Y\)
and the lepton number symmetry U(1)\(_L\) breaks down
spontaneously together with the breaking of  G.
The Majoron, the Nambu-Goldstone (NG) boson associated with
the U(1)\(_L\) breaking, is induced as a linear combination of
only doublet Higgs bosons which is of course not the case
in the singlet Majoron (SM) model \cite{chi}
or the triplet Majoron (TM) model \cite{gel}.
In this respect, we refer to the present  model
as the doublet Majoron (DM) model.
There is also a model in which the Majoron belongs to the doublet
representation \cite{ber1}, but that version is now ruled out
experimentally by the nonobservation of Z$^0 \rightarrow$ invisible
light scalars.
In  the present model the light neutrino masses come about through
the seesaw mechanism \cite{yan1}, while in the model of Ref.~\cite{ber1}
they  are induced radiatively.

One motivation of Majoron models in general is that they may
induce neutrino decay through  Majoron emission \cite{chi}.
This decay can  provide a way to evade the cosmological constraint on the
neutrino mass \(m_\nu\), that is, \(m_\nu < \) 100  eV  or
\(m_\nu > \) a few GeV for a stable neutrino.
In fact, this decay in both the TM and SM models is strongly suppressed
and it turns out to be almost cosmologically irrelevant:
in the TM model the decay  only takes place through a loop diagram \cite{geo},
while in the SM model it is  suppressed at tree level
by the  fourth power of the seesaw factor
\( (m_\nu / M ) \) (\(M\) is the mass scale for
the right-handed gauge-singlet neutrino) \cite{sch}.
In the DM model, on the other hand, the decay occurs
by the same mechanism as in the SM model,
but the suppression of the seesaw factor is weaker; the power reduces to
three instead of four.
This means that neutrino decay in the DM model could well be
cosmologically relevant.

Some time ago, the possibility of a  \(\tau\) neutrino of about 10 MeV
in mass was proposed in a scenario of baryogenesis at the electroweak
phase transition in the SM model \cite{coh}.
However it appears difficult to have such
a mass range which also satisfies the cosmological constraint
unless one chooses an unnaturally small \(M\).
We aim to investigate the decay of such a neutrino more closely for
a similar mass range using the DM model.
We will also take into account the constraint of Majoron  emission from
the supernova 1987A \cite{cho} and that of nucleosynthesis \cite{ber2,wal}.

There is another motivation for looking into Majoron models.
Although the seesaw mechanism is an attractive idea to account for the mass
hierarchy of the neutrinos compared to the charged leptons \cite{yan1},
once it is combined with the observation that the anomalous
baryon and lepton number \(B + L\) violating process
is in thermal equilibrium at an early stage of the
Universe \cite{kuz}, it leads to a very strong bound on the neutrino masses
\cite{fuk,har}.
If both the \(B+L\) violation from the anomalous process and the
\(L\) violation from the gauge-singlet neutrino mass term are active
together at a certain time during the Universe's evolution,
any primordial excess in \(B\) or \(L\) is wiped out.
To avoid this situation, a neutrino mass must be
less than 1 eV as long as the anomalous process is fast enough up to
a temperature of about \(10^{12}\) GeV  \cite{pec}.
This makes any neutrino an unsuitable candidate for a component of
dark matter.

One way to evade the above constraint is to make U(1)\(_L\)
an exact symmetry above  the electroweak phase transition.
In the DM model, the \(L\) symmetry is exact before spontaneous
symmetry breaking even though
the gauge-singlet neutral fermion has mass, and it breaks down
at the electroweak scale, which is a natural
consequence of its Higgs representation.
Provided that this phase transition is first-order and that
the anomalous process
is suppressed after it \cite{tuk},
the \(L\) and \(B+L\) violating processes cannot coexist above
or below the phase transition.

\paragraph{2. The model}~

We now describe the DM model.
A basic observation that characterizes the DM model is that a Majorana mass
for the gauge-singlet neutral fermions \(N_a\) (\(a= 1, 2, 3\)
denotes generations)
does not immediately imply \(L\) violation.
Since \(N_a\)
belong to a different representation from the doublet leptons \(L_a\),
they can have a lepton number different from that of \(L_a\).
We assign zero lepton  number to  \(N_a\) so that their Majorana mass
terms do not violate \(L\).
Instead, we introduce a  Higgs doublet \(H_1\),
different from the ordinary one
\(H_0\), and assign \(L = -1\) to it. (The hypercharge \(Y\) is the same for
all the doublet Higgs bosons, \(Y  = 1\).)
The Yukawa coupling between \(L_a\) and \(N_a\) is provided  by \(H_1\).
The spontaneous breakdown of G, as \(H_0\) and \(H_1\) get
vacuum expectation values,
is accompanied by the  violation of U(1)\(_L\).

Actually, this simple extension of the minimal standard model to a Majoron
model turns out not to be enough.
The Majoron obtained this way has components in both \(H_0\) and \(H_1\).
As a result, the Majoron couples to the charged leptons through
the hypercharge current \(j_\mu^Y\) as well as through the lepton current
\(j_\mu^L\).
This coupling allows Majoron emission in  Compton scattering
and significantly contributes to the stellar cooling of red giants
\cite{geo,dic}.
To avoid this coupling, we introduce another Higgs doublet
\(H_2\) and assign \(L = 1\) to it.
As we will see below, we can suppress the Majoron coupling to the charged
leptons by a cancellation between \(H_1\) and \(H_2\).

\newcommand{\e}{{\mbox{\scriptsize e}}}
\newcommand{\nue}{{\nu_e}}
\newcommand{\numu}{{\nu_\mu}}
\newcommand{\nutau}{{\nu_\tau}}
\newcommand{\T}{{\mbox{\scriptsize T}}}
The Lagrangian is the same as in the standard model except for the part
including the extra Higgs bosons and gauge-singlet neutral fermions.
The corresponding part is
\bea
{\cal L} & = & \sum_{\alpha } |D_\mu H_\alpha |^2 - V( H_\alpha)
\nonumber\\
&&+\sum_a\left[  i L^\dagger_a (D_\mu \bar\sigma^\mu) L_a
+ i C_a^\dagger (D_\mu \bar\sigma^\mu) C_a  + i N_a^\dagger( \partial_\mu
\bar\sigma^\mu) N_a \right] \nonumber\\
&& + \sum_{a,b} \left[ H_0^\dagger \Gamma^0_{ab} ( L_a^\T i\sigma^2 C_b )
+ \tilde H_1^\dagger \Gamma^1_{ab} (L_a^\T i\sigma^2 N_b )
+ {1\over 2} M_{ab} (N_a^\T i\sigma^2N_b )  + (\mbox{h.c.}) \right],
\eea
where all the lepton fields have been written in  left-handed two-component
notation; \(L_a\) denotes  the doublet leptons
\be
L_a = \left( \begin{array}{c} \nu_{\e} \\ \mbox{e} \end{array}\right),
\left( \begin{array}{c} \nu_\mu \\ \mu \end{array}\right) ,
 \left( \begin{array}{c} \nu_\tau \\ \tau \end{array}\right);
\ee
\( C_a \) the singlet charged leptons \(\mbox{e}^c\), \(\nu^c\), and
\(\tau^c\); \(N_a\) the singlet neutral fermions;
\(\tilde H = i\tau^2 H^*\);
\(D_\mu \) is the covariant derivative and
\(\bar\sigma^\mu = ( 1, - \vec\sigma ) \);
Note that \(H_2\) cannot have a Yukawa coupling because of its quantum
number \(Y = 1\) and \( L = 1 \).

The Higgs potential which is invariant under G
and U(1)\(_L\) consists of thirteen terms, i.e.,
three of the type \( (H_\alpha^\dagger H_\alpha) \),
three of the type \( (H_\alpha^\dagger H_\alpha)^2 \),
three of the type \( ( H_\alpha^\dagger H_\alpha )
( H_\beta^\dagger H_\beta ) \),
three of the type \( (H_\alpha^\dagger H_\beta) (H_\beta^\dagger H_\alpha ) \),
and \( (H_0^\dagger H_1)(H_0^\dagger H_2) \) plus
its hermitian conjugate.
For the last term, we have used the rephasing of \(H_\alpha\) to absorb
a possible complex phase factor.
Thus \(V\) has thirteen {\em real} parameters and
has no explicit CP violation.
For a wide range of parameters, \(V\) breaks  G\(\times\)U(1)\(_L\) down
to  U(1)\(_{\mbox{\scriptsize em}}\) spontaneously
with  nontrivial {\em real} expectation values \(v_\alpha\) of \(H_\alpha\).
Let us write the scalar fields as
\be H_\alpha = \left( \begin{array}{c} h_\alpha \\ \Phi_\alpha
\end{array}\right),  \qquad
\Phi_\alpha = e^{i (\varphi_\alpha / \sqrt 2 v_\alpha) }
\left( v_\alpha + {1\over \sqrt 2}  \phi_\alpha \right).
\label{Phi}
\ee

These thirteen parameters are phenomenologically constrained for
the following reason.
In the imaginary parts of the neutral sector of the Higgs bosons,
there are two NG bosons
associated with  the violation of U(1)\(_Y\) and U(1)\(_L\).
An appropriate linear combination is the one absorbed into Z\(^0\), and the
other orthogonal combination remains massless and becomes the Majoron.
The specific form of the linear combination for the Majoron in terms of
\(\varphi_\alpha\) is obtained by working with the currents
\(j_\mu^L\) and  \(j_\mu^Y\),
\be
j^{L\,\mu} = i \sum_\alpha L (H_\alpha)
\{ H_\alpha^\dagger D^\mu H_\alpha - (D^\mu
H_\alpha )^\dagger H_\alpha \} + \sum_f L(f) ( f^\dagger \bar\sigma_\mu f )
,\label{djl}
\ee
where \(L(H_\alpha)\) and  \(L(f)\) denote the lepton number assignment for
 \(H_\alpha\) and the fermion species \(f\);
\(j_\mu^Y\) is given by the same equation with \(Y(H_\alpha)\) and
\(Y(f)\), the hypercharges, instead of \(L(H_\alpha)\) and \(L(f)\).
We write the current-conservation equations
using  Eq.~(\ref{Phi}).
They are
\bea
\partial_\mu\partial^\mu \sum_\alpha L(H_\alpha) v_\alpha  \varphi_\alpha
& = &
- {g \over \sqrt 2 \cos\theta_W}
\left( \sum_\alpha L(H_\alpha) v_\alpha^2\right)
 \partial_\mu Z^\mu +  {1\over \sqrt 2} \partial_\mu j^{L\,\mu} (\mbox{f})
+ ...\label{jl}\\
\partial_\mu\partial^\mu \sum_\alpha v_\alpha  \varphi_\alpha
& = &
- {g \over \sqrt 2 \cos\theta_W}
\left( \sum_\alpha v_\alpha^2 \right)
\partial_\mu Z^\mu +  {1\over \sqrt 2} \partial_\mu
 j^{Y\,\mu} (\mbox{f}) + ...,   \label{jy}
\eea
where  \(\theta_W\) is the
Weinberg angle and \(j^{L(Y)}_\mu(\mbox{f})\)
denotes  the fermion component in each current.
These equations describe the motion of the NG bosons; the right-hand sides
represent the interactions with the other fields.
The Majoron is the combination that does not have the linear \(Z_\mu\) term
in its equation.
We can thus simply use Eqs.~(\ref{jl}) and (\ref{jy})
to extract out the equation for the Majoron.
This procedure reveals that the Majoron has a coupling of the
form \( g_{\varphi\e\e} \partial_\mu\varphi  j^{Y\,\mu} \)
with the strength \(g_ {\varphi\e\e} \) proportional to
\be \sum_\alpha L(H_\alpha) v_\alpha^2.  \label{CC}\ee
This coupling is highly constrained to be less than about
$10^{-13}$ divided by the electron mass
from the astrophysical consideration
already mentioned \cite{geo,dic}.

The reason we employed \(H_2\) is to cancel the \(H_1\) contribution
in (\ref{CC}) \cite{val}.
Note that it vanishes for \(L(H_1)\) =  \(- L(H_2)\) and
\(v_1 = v_2\).
Hence we parametrize \(V\) as
\bea
V & =&  \sum_\alpha \lambda_\alpha
( H_\alpha^\dagger H_\alpha - v_\alpha^2 )^2
+   \sum_{\alpha<\beta} \eta_{\alpha\beta}
 ( H_\alpha^\dagger H_\alpha - v_\alpha^2 )
( H_\beta^\dagger H_\beta - v_\beta^2 ) \nonumber \\
&&  + \sum_{\alpha<\beta} \zeta_{\alpha\beta}
\left[  ( H_\alpha^\dagger H_\alpha )( H_\beta^\dagger H_\beta )
- ( H_\alpha^\dagger H_\beta)( H_\beta^\dagger H_\alpha) \right] \nonumber \\
&& + \xi \left[  ( H_0^\dagger H_0 )( H_1^\dagger H_1 ) +
 ( H_0^\dagger H_0 )( H_2^\dagger H_2)  - (H_0^\dagger H_1)(H_0^\dagger H_2)
- (H_1^\dagger H_0)(H_2^\dagger H_0)
\right],
\eea
and further require it to be symmetric under the exchange
\( H_1 \leftrightarrow H_2 \). The parameters are then constrained:
\( v_1 = v_2 \equiv v_L \), \(\lambda_1 = \lambda_2 \equiv \lambda\),
\(\eta_{01} = \eta_{02} \equiv \eta \),
and \(\zeta_{01} = \zeta_{02} \equiv \zeta \).
\(V\) has a minimum at the same expectation value for \(\Phi_1\) and
\( \Phi_2 \) and the Majoron coupling to \(j_\mu^Y\) vanishes.

Since the couplings to fermions are different for \(H_1\) and \(H_2\),
the above exchange symmetry is no longer exact once fermion-loop
corrections are included.  If we adopt the ``effective-potential'' method
for evaluating
the  quantum corrections to \(V\), they are typically of order \(\gamma^4\)
(\(\gamma\) is the Yukawa coupling of \(H_1\) to the neutral fermions).
We assume these corrections are fine-tuned so that they will not have a
significant contribution to the \(j_\mu^Y\) coupling.

We now find out  explicitly the particle spectrum in the Higgs sector.
For the charged scalars \(h_\alpha\), the mass matrix is given by
\begin{equation}
\left(\begin{array}{ccc}
2 ( \zeta + \xi ) v_L^2 & -( \zeta + \xi ) v_0 v_L & - (\zeta + \xi ) v_0 v_L
\\
-( \zeta + \xi ) v_0 v_L &  ( \zeta + \xi ) v_0^2 + \zeta_{12} v_L^2 &
- \zeta_{12} v_L^2 \\
-( \zeta + \xi ) v_0 v_L & - \zeta_{12} v_L^2 &  ( \zeta + \xi ) v_0^2 +
\zeta_{12} v_L^2
\end{array}\right).
\end{equation}
The mass eigenstates and their masses are readily evaluated:
\bea
h_G &=& \cos \beta h_0 + {\sin \beta\over \sqrt 2 } (h_1 + h_2); \quad m^2 =0,
\\
h_L &=& {1\over \sqrt 2} (h_1 - h_2); \quad m^2 =
(\zeta + \xi ) v_0^2 + 2 \zeta_{12} v_L^2,
\\
h_H & = & - \sin \beta h_0 + {\cos \beta \over \sqrt 2 } (h_1 + h_2) ;
\quad m^2 = (\zeta + \xi ) ( v_0^2 + 2 v_L^2 ), \eea
where the angle \(\beta\) is defined by
\be \tan \beta = {\sqrt 2 v_L \over v_0 }. \ee
The zero eigenstates \(h_G\) correspond to the NG bosons associated with the
SU(2) breaking and they are absorbed into the W\(^\pm\) gauge bosons.
For the real part of the neutral sector,
the mass matrix is
\begin{equation}
\left(\begin{array}{ccc}
4 \lambda_0 v_0^2 & 2 \eta v_0 v_L & 2 \eta v_0 v_L \\
2 \eta v_0 v_L & 4 \lambda v_L^2 + \xi v_0^2 & 2\eta_{12} v_L^2 - \xi
v_0^2 \\
2 \eta v_0 v_L  & 2\eta_{12} v_L^2 - \xi v_0^2
&  4 \lambda v_L^2 + \xi v_0^2
\end{array}\right),
\end{equation}
and the mass eigenstates and masses are
\bea
\phi_L & = & {1\over \sqrt 2} (\phi_1 - \phi_2); \quad
m^2 = (4\lambda - 2 \eta_{12}) v_L^2 + 2 \xi v_0^2, \\
\phi_+ & = & \cos\alpha\, \phi_0 + {\sin \alpha \over \sqrt 2}
(\phi_1 + \phi_2);
\quad m^2 = 4\lambda_0 v_0^2 + \tan \alpha\, {4\over\sqrt 2} \eta v_0 v_L, \\
\phi_- & = & -\sin\alpha\, \phi_0 + {\cos \alpha \over \sqrt 2}
(\phi_1 + \phi_2);
\quad m^2 = (4\lambda + 2 \eta_{12}) v_L^2 -\tan \alpha\, {4\over \sqrt2}
\eta v_0 v_L ,
\eea
where \(\alpha\) is defined by
\be \cot \alpha -\tan \alpha =
{\sqrt 2 \over  \eta } \left[ \lambda_0 { v_0\over v_L} -
\left(\lambda + {\eta_{12}\over 2}\right) {v_L\over v_0} \right],
\quad -{\pi\over 4} < \alpha < {\pi\over 4} .\ee
Similarly for the imaginary part, the mass matrix is
\begin{equation}
\xi \left( \begin{array}{ccc}
4  v_L^2 & - 2 v_0 v_L & - 2 v_0 v_L \\
- 2 v_0 v_L & v_0^2 & v_0^2 \\
- 2 v_0 v_L & v_0^2 &  v_0^2
\end{array} \right).
\end{equation}
The mass eigenstates are
\bea \varphi_A & =&   -\sin \beta \varphi_0 + {\cos \beta \over \sqrt 2}
(\varphi_1 + \varphi_2 ); \quad m^2 = 2 \xi ( v_0^2 + 2 v_L^2 ), \\
\varphi_L & = & {1\over \sqrt 2} (\varphi_1 - \varphi_2 );
\quad m^2 = 0 \\
\varphi_G & = &
\cos \beta \varphi_0 + {\sin \beta \over \sqrt 2 } (\varphi_1 +
\varphi_2 ); \quad m^2 = 0.
\label{vpG}
\eea
The combination \(\varphi_G\) is absorbed into the Z\(^0\) gauge boson and
\(\varphi_L\) is the Majoron.

The motion of the Majoron, especially its interaction with the other fields,
is now solely described by the \(j_\mu^L\) conservation, Eq.~(\ref{jl}).
We obtain the effective interaction Lagrangian
\( {\cal L}_{\mbox{\scriptsize eff}} \) by requiring that the resulting
Euler-Lagrange equation with respect to \(\varphi_L\) coincides with
Eq.~(\ref{jl}). Up to cubic terms, we get
\bea
{\cal L}_{\mbox{\scriptsize eff}} & = &
\left( { \partial^\mu \varphi_L \over v_L } \right)
\left[
{1\over 2} j_\mu^{L}(\mbox{f})
+ {g v_L \over\cos\theta_W}  Z_\mu \phi_L
- g v_L \left( W_\mu^+  h_L^\dagger + W_\mu^-  h_L \right)  \right.
\nonumber\\
&& + {1\over 2} \left( \partial_\mu \varphi_L \right) \left( \cos\alpha\,\phi_-
+ \sin\alpha\,\phi_+ \right)
+ \cos\beta\left(\partial_\mu\varphi_A\right) \phi_L
\nonumber \\
&& \left.
+ \left({i\over 2}\right) \cos\beta
\left\{ ( \partial_\mu h_H^\dagger ) h_L
               - h_H^\dagger ( \partial_\mu h_L )
       + ( \partial_\mu h_L^\dagger ) h_H
               - h_L^\dagger ( \partial_\mu h_H ) \right\}
 \right]. \label{Leff}
\eea
Reflecting on the NG-boson nature of \(\varphi_L\), we have written
its interaction in
the derivative-coupling form.
The first term in the bracket includes the coupling to the neutrinos and
induces neutrino decay.
The second term is important since it describes the coupling to Z\(^0\).
Remember that the mass of the accompanying  scalar particle, \(\phi_L\), is
\((4\lambda -2\eta_{12})v_L^2 + 2 \xi v_0^2 \).
The values for \(v_0\) and \(v_L\) are constrained
by \(\sqrt{v_0^2 + 2 v_L^2} = \) 174 GeV, but otherwise they are
free.
The expression for the \(\phi_L\) mass can naturally give a bigger value than
the Z\(^0\) mass.
Hence the decay \( Z^0  \rightarrow \varphi_L\, \phi_L\) can be forbidden
kinematically or else this model
would be ruled out by the present LEP experiments.
Note also that in the DM model all the Higgs bosons belong to the doublet
representation and \(v_L\) and \(v_0\) are
free from the \(\rho\)-parameter constraint.
\newpage
\newcommand{\jlfer}{\mbox{\(j_\mu^L(\mbox{f})\)}}
\newcommand{\D}{{\mbox{\scriptsize D}}}
\paragraph{3. Neutrino decay}~

We now look into the neutrino decay
\be \nu_\tau \rightarrow \nu_\mu\, \varphi_L \quad \mbox{or} \quad
\nu_{\e}\, \varphi_L. \ee
They are induced by the first term in the bracket in Eq.~(\ref{Leff}).
The existence of these flavor-changing processes can be seen in two steps.
First, the conservation of the neutrino portion of \jlfer\ is violated in the
symmetry-broken phase by the appearance together of both the Dirac mass,
\be m_{\D\,ab} = v_L \Gamma^1_{ab}, \ee
and the Majorana mass \(M_{ab}\).
This explains why the Majoron coupling to a charged lepton
is suppressed even though \jlfer\
has charged-lepton components: their masses  are necessarily of
the Dirac type, hence that part of \jlfer\ is automatically conserved
at tree level.
Second, although \jlfer\ is diagonal with respect to \(\nu_a\),
the mass diagonalization procedure involves both \(\nu_a\) ($L=1$)
and \(N_a\) ($L=0$),
and this  generates nondiagonal flavor-changing vertices.
This is in contrast to the TM model, where no singlet neutrino is
involved and the current is flavor-diagonal even after the
diagonalization of the Majorana mass matrix for \(\nu\).

Let us see this second point in detail for the  seesaw
mass matrix and obtain the neutrino decay  vertex explicitly.
In the symmetry-broken phase the physical fields are the mass
eigenstates.  We write the  neutrino fields in  \jlfer\ in terms of
these. It is done by the following replacement:
\be \left(  \begin{array}{c} \nu \\ N \end{array} \right) \rightarrow
U  \left(  \begin{array}{c} \nu \\ N \end{array} \right) =
\left(  \begin{array}{cc} U_1 & U_2 \\ U_3 & U_4  \end{array} \right)
\left(  \begin{array}{c} \nu \\ N \end{array} \right), \ee
where we have suppressed the generation index; \(U_i\) are
3 \(\times\) 3 matrices.
The unitary matrix \(U\) is obtained  by
\be
U^\T \left(  \begin{array}{cc} 0 & m_\D \\ m_\D^\T & M \end{array} \right) U
= \mbox{diag}(m_{\nue}, m_{\numu}, m_{\nutau}, m_{N_1}, m_{N_2}, m_{N_3}).
\ee
Although \(U\) is unitary, \(U_1\) is not necessarily so.
Thus the light neutrino components in \jlfer\ after the replacement,
\(\nu^\dagger U_1^\dagger U_1 \bar\sigma^\mu \nu \),  are
not diagonal in general and allow the flavor-changing decay.
For a typical seesaw mass matrix \((m_\D \ll M) \),
\(U_2\) and \(U_3\) are of order \(m_\D/M\) and  the deviation of
\(U_1\)  from a unitary matrix is small, of order \((m_\D/M)^2\).
The matrices \(U_i\) have been obtained order by order
in \(m_\D/M\) \cite{sch}. \(U_1\) is  given explicitly up to  second
order as
\be U_1 = \left( 1 - {1\over 2} m_\D^*{ 1\over M^2 } m_D^\T  \right) V,
\ee
where the unitary matrix \( V  \) is defined by
\be V^\T \left( - m_\D {1\over M} m_\D^\T \right) V
= \mbox{diag} ( m_{\nue}, m_{\numu},
m_{\nutau} ),
\ee
and in solving \(U_1\) we have assumed without loss of generality
that \(M_{ab}\) is already diagonal
with positive eigenvalues.
The eigenvalue \(m_\nu\) has the typical seesaw size, \(m_\D^2/ M \).
The neutrino decay vertex is now given by
\be {\cal L} = {\varphi_L\over 2 v_L} \sum_{a,b}
 {\partial_\mu } R_{ab} (\nu_a^\dagger
\bar \sigma ^\mu \nu_b ), \quad R \equiv V^\dagger m_\D^* {1\over M^2}
m_\D^\T V \label{ld}\ee
and the flavor-changing mixing is of order \(m_\nu/ M\).

The decay width of \(\nu_\tau\) is readily evaluated.
For \(m_\nutau \gg m_\numu, m_\nue\), we obtain
\be \Gamma = \sum_{a = \mu,\e}  {1  \over 64\pi }
|R_{\nutau \nu_a}|^2 { m_\nutau^3  \over v_L^2 }
 \simeq  { \gamma^2 \sin^2 \theta \over 64\pi }
\left( {m_{\nutau} \over M } \right)^3 m_{\nutau}, \label{Gam}\ee
where we  parametrize \(|R|\) as \((m_{\nutau} \sin \theta / M) \)
with a mixing angle \(\theta\) and \( ( m_{\nutau}/ v_L )^2 \) as
\(( \gamma^2 m_{\nutau} /M ) \) with the Yukawa coupling \(\gamma\)
for the \(\tau\) neutrino.

In the SM model this decay is also similarly described
but the power of \( m_{\nutau} / M\)
in the final result (\ref{Gam})
is 4 instead of 3. This is because the corresponding scalar expectation value
\(v_L\) in the SM model is related to the mass of the gauge-singlet neutrinos.

\newcommand{\Omeganu}{\Omega_{\nu\bar\nu}}
\newcommand{\Omegagamma}{\Omega_{\gamma\nu}}

Let us compare this width with the cosmological bound.
The thermal history of neutrinos in the expansion of the Universe is well
studied\footnote{
The key process to get the relic energy density of a neutrino species
in Ref.~\cite{kol} is \(\nu\,\bar\nu \leftrightarrow f\,\bar f\).
In the following, we assume that extra processes,
intrinsic in a Majoron model such as \(\nu\,\nu \leftrightarrow
\varphi_L\,\varphi_L\), will not change the results drastically.}
\cite{kol}.
Their relic density parameter \(\Omeganu \) is given by
\be \Omeganu h^2 \sim {m_\nu\over 91.5 \mbox{eV}} \ee
for each light neutrino, {\it i.e.} that which is
relativistic at its decoupling temperature, or by
\be \Omeganu h^2 \sim \left({m_\nu\over \mbox{GeV}}\right)^{-2} \label{hv1}\ee
for each heavy neutrino, {\it i.e.} that which is nonrelativistic at
its decoupling temperature. [The notation is the same as in
Ref.~\cite{kol}.
Since we are interested in a narrow mass range, around 10 MeV, we have
neglected a logarithmic dependence on \(m_\nu\) in Eq.~(\ref{hv1}).]
Thus a neutrino species that is heavier than 100 eV or lighter than
a few GeV must decay.
For the lifetime constraint we use here
that which comes from a consideration on
``structure formation'' \cite{ste}, which gives a much stronger bound than
the condition of not overclosing the Universe by the relic density
of the decay products \cite{dis}.
This requires the relic density parameter of the decay product
to be smaller than that of the radiation, {\it i.e.}
the photon and the light neutrinos,
\be  \Omegagamma h^2 \sim 4 \times 10^{-5}. \ee
The relic density of the relativistic decay product decreases faster than
that of the nonrelativistic matter and is approximately
given by \(\Omeganu R_{\D}\), where \(R_{\D}\) is a scale parameter
related to the lifetime \(\Gamma^{-1}\)  by
\be \Gamma^{-1} \sim {4.9 \times 10^9 R_\D^2 \over (\Omegagamma h^2)^{1/2} }
\; \mbox{year}. \ee
Thus we get
\bea \Gamma^{-1} &< &
3 \times 10^2 \left( m_\nu\over 1 \mbox{MeV} \right)^{-2}
\mbox{sec} \quad\mbox{ for light neutrino},\label{lg}\\
\Gamma^{-1} &<  &
4 \times 10^{-2} \left( m_\nu\over 1 \mbox{MeV} \right)^4
\mbox{sec} \quad\mbox{ for heavy neutrino}.   \label{hv}
\eea

We demonstrate in Fig.~1 that it is indeed possible for \(\nu_\tau\) to
be as heavy as 10 MeV for a choice of parameters \(\gamma^2 = 10^{-2}\)
and \(\sin^2 \theta = 10 ^{-2} \)  as an example.
We depict the allowed area in the \((m_{\nutau}\)--\(M)\) plane.
The lines AB and BC come from the cosmological bound for
light and heavy neutrino, respectively. They are obtained by
Eq.~(\ref{lg}) or (\ref{hv}) with (\ref{Gam}).
We have also taken into account a constraint coming from
supernova cooling by Majoron emission \cite{cho}, which is shown by
the line CD.
The region allowed by the cosmological constraint mostly corresponds
to that of ``Majoron trapping'' and it gives \cite{cho}
\be \left( m_\nu\over 1 \mbox{MeV}\right)\left({ 1 \mbox{GeV}\over
v_L}\right)^2
> 3.3 \times 10^{-3}.\ee
Note that the values for \(m_{\nutau}\) and \(M\) implicitly determine the
value of \(v_L\) by the relation \( \gamma v_L \sim \sqrt{ M m_{\nutau} } \).
There is of course also the laboratory upper bound of 35 MeV
on \(m_{\nutau}\).

The values of \(v_L\) for the allowed region in Fig.~1 are relatively
low compared with \(v_0 \sim \) 170 GeV; they are typically a factor of 5
or so less.
This predicts a relatively light Higgs boson, the \(\phi_-\) of Eq. (17).
Since this Higgs boson is orthogonal to \(\phi_L\) which couples
the Majoron \(\varphi_L\) to the Z\(^0\) boson,
there is no conflict with the experimental data for the Z\(^0\) width.
It can, however, be a rare decay product of Z\(^0\) in
\be \mbox{Z}^0 \rightarrow \phi_-\, f\,\bar f. \label{pc}\ee
The branching fraction of this process is suppressed by
\be (- \cos \beta \sin \alpha + \sin\beta \cos\alpha )^2 \sim 10^{-2} \ee
compared with that of the single Higgs boson of the minimal standard model
and thus \(\phi_-\) can still have a mass below the latter's experimental
lower bound of about 60 GeV.
Through its \(\phi_0\) component, \(\phi_-\) decays into visible channels
such as charged fermion pairs,
and may thus be observed in future Higgs-boson search experiments.
We have drawn the line AE that corresponds to \(v_L\) = 10 GeV
in Fig.~1; \(\phi_-\) with this mass has  roughly the same
branching fraction as the standard Higgs boson of 60 GeV
in the process
(\ref{pc}) \cite{gun}.

Finally we consider the effect of nucleosynthesis.
The number of relativistic degrees of freedom at the temperature of
about 1 MeV strongly affects the abundance of light elements \cite{kol}.
Using the best current nuclear physics data and astronomical observations,
the number in terms of corresponding light neutrino species,
\(N_\nu\), is restricted to be less than 3.3 \cite{wal}.
The Majoron contribution to \(N_\nu\) has been considered extensively
in Ref.~\cite{ber2}.
The key process for the Majoron to be in thermal contact
with the neutrinos is \(\nu\nu \leftrightarrow \varphi_L \varphi_L\).
Its decoupling temperature \(T_\D\) is estimated  by comparing
the inverse mean free path of
the process with the expansion rate of the universe,
\be T_\D \sim 10^3 \sqrt{ g_*(T_\D) } { v_L^4 \over m_{\nu}^2 }
{ 1\over m_{\mbox{\scriptsize Pl}} }. \ee
The decoupling temperature from  the
\(\tau\) neutrino is the lowest among the
three neutrinos and it is
much lower than 1 MeV  for the parameters that are allowed in Fig.~1.
The Majoron keeps thermal equilibrium as long as the \(\tau\) neutrino does.
Thus its decoupling temperature cannot be high enough to suppress its
contribution to \(N_\nu\) \cite{ber2},
which is given by (8/7)(1/2)~\(\sim\)~0.6.

To keep \(N_\nu\) within the above-mentioned bound,
the \(\tau\) neutrino  needs to be  nonrelativistic at its decoupling
temperature and its
energy density must be small compared to that of the radiation.
Since the enery density of  massive matter after  decoupling decreases
at a slower rate  than that of the massless degrees of freedom,
it may significantly contribute to the total energy density again afterwards.
If this matter domination takes place at the time of nucleosynthesis,
it would affect the primordial abundance of light elements \cite{kol2}.
The cosmological constraint on  the \(\tau\) neutrino lifetime we have used
does not allow its energy density or that of its decay products
to exceed  that of the light degrees of freedom.
But it may still contribute partially to \(N_\nu\).
Based on a detailed study of the
\(\tau\) neutrino lifetime and mass constraints from nucleosynthesis,
we thus further require its lifetime to be shorter than 1 second
\cite{kol2,kaw}.
We show the corresponding boundary by a dashed line in Fig.~1,
the region to the right of which remains allowed.
Under the  conditions that \(\tau\) neutrino is strongly nonrelativistic
at the decoupling temperature and it decays fast enough before
primordial nucleosynthesis,
\(N_\nu\) is given by the sum of the contributions from
the Majoron and the two lighter neutrinos, i.e. 2.6.

In summary, we have presented a new Majoron model.
It allows the \(\tau\) neutrino to be in the 10-MeV mass range.
\vspace{0.3in}
\begin{center} {ACKNOWLEDGEMENT}
\end{center}

This work was supported in part by the U.S. Department of Energy under
Contract No. DE-AT03-87ER40327.

\newpage
\begin{center}
\bf Figure caption
\end{center}
\begin{description}
\item[Fig.~1] Region of the \((m_{\nutau}\)--\(M)\) plane
allowed as the result of various constraints for $\gamma^2 = \sin^2 \theta
= 10^{-2}$.
\end{description}

\end{document}